\documentclass[12pt]{article}
\usepackage{graphicx}
\usepackage{epsfig}

\usepackage{a4wide,amsmath,amssymb}
\long\def\symbolfootnote[#1]#2{\begingroup%
\def\thefootnote{\fnsymbol{footnote}}\footnote[#1]{#2}\endgroup}

\begin{document}
% declarations for front matter
%\begin{titlepage}
\rightline{UG-FT-172/04}
\rightline{CAFPE-42/04}
\rightline{hep-ph/0410082}
\vspace{.5cm}

\begin{center}
{\large
\textbf{Discrete regularisation of localised kinetic
  terms}}\symbolfootnote[2]{Presented at the 7th DESY Workshop on
  Elementary Particle Theory ``Loops and Legs in Quantum Field
  Theory'', Zinnowitz, April 25-30, 2004.}\symbolfootnote[1]{This  
       work has been supported in part by MEC and FEDER 
       Grant No. FPA2003-09298-C02-01,
       by Junta de Andaluc{\'\i}a Group FQM 101, by the %European 
       %Community
EC's Human Potential Programme under contract 
       HPRN-CT-2000-00149 Physics at Colliders, and by PPARC.}
\vspace{.5cm}

\textbf{F. del Aguila$^a$, M. P\'erez-Victoria$^{a,b}$ and J. Santiago$^c$}

\vspace{.5cm}
\textit{$^a$Departamento de F{\'\i}sica 
Te\'orica y del Cosmos and CAFPE, \\ Universidad de Granada,  
                E-18071 Granada, Spain}
\vspace{.25cm}

\textit{$^b$Dipartimento di Fisica ``G. Galilei'', 
                Universit\`a di Padova and \\ INFN, Sezzione di Padova, 
                Via Marzolo 8, I-35131 Padua, Italy} 
\vspace{.25cm}

\textit{$^c$IPPP,%nstitute for Particle Physics Phenomenology, 
                University of Durham, %\\
                South Road, Durham DH1 3LE, United Kingdom}

\end{center}

\centerline{{\large \textbf{Abstract}}} 
\vspace{.5cm}
      
We investigate the behaviour of 5d models with general brane kinetic terms by 
discretising the extra dimension.We show that in the continuum limit the 
Kaluza-Klein masses and wave functions are in general nonanalytic in the 
coefficients of brane terms.
\vspace{1pc}

%\end{titlepage}

% typeset front matter (including abstract)

\section{INTRODUCTION}

Field theories with extra dimensions give a new perspective on 
many important issues in particle physics, such as the hierarchy problem, 
(super)symmetry breaking or the structure of flavour (for recent reviews 
see \cite{F,C,HM}). 
In general they must be understood as effective 
theories valid below some scale, above which a more fundamental 
theory must be at work. 
It is then important to keep in mind the presence of higher order terms in 
the low-energy expansion of the effective lagrangian. The situation is 
particularly interesting when the theory contains lower dimensional defects, 
called ``branes'' from now on. In this case one expects localised corrections 
to the lowest-dimensional terms. In fact, infinite localised corrections are 
induced by quantum corrections when the bulk fields couple to brane 
fields~\cite{DGP} and in orbifold compactifications~\cite{GGH01,CMS}. The 
divergences must be cancelled by localised counterterms, and the coefficients 
(couplings) of the corresponding brane terms run with the scale, so that they 
cannot be chosen to vanish at all scales. 

In particular, the effective action of generic brane models contains
brane kinetic terms (BKT). These terms modify the phenomenological
predictions of the model~\cite{CTW,APS03a} and have been studied in
different contexts~\cite{DGP,DHR} (for a list of references see
\cite{APS03c}). In~\cite{APS03b} we have studied the implications of
the most general BKT for particles of spin 0, $\frac{1}{2}$ and 1 in
five dimensions, with the fifth dimension compactified on an
orbifold. Some of the allowed BKT have dramatic effects in the
spectrum of Kaluza-Klein (KK) masses and wave functions. To understand
this one should keep in mind that, in general, the equations of motion
(or, equivalently, the differential equations for the KK reduction)
are not well-defined in the presence of the delta functions which
impose localisation in the BKT. A sensible prescription is 
to regulate the delta functions using thick branes, perform the KK
reduction and then take the thin brane limit.
However, it turns out that the limits of thin
brane and vanishing BKT do not always commute. This signals a
breakdown of the perturbative expansion of the effective lagrangian,
raising the question of the consistency of the
approach.

Inspired by deconstructed theories, we study here a discretised
version of five dimensional models with BKT terms, for fields of spin
$0, \frac{1}{2}$ and $1$. We take the continuum limit and compare with
the corresponding results in (the thin brane limit of) the thick brane
regularisation.  
This lattice regularisation is complementary to the thick brane
regularisation, in the sense that it is the bulk which is regularised
rather than the brane. From a practical point of view, it turns out
that discretisation gives a better handle of analytic computations
than thick brane regularisation. 
%Even though we shall be primarily
%interested in the continuum limit here, we stress that one should keep
%a finite spacing if the deconstructed theory is regarded as a
%microscopic theory completing the five dimensional model in the
%continuum.   
In~\cite{aliphatic} we take this point of view seriously and study a
full deconstructed model. In particular we calculate loop corrections
and show that BKT are also generated in these models if the discrete
analogue of Poincar\'e invariance is broken at some point. This
communication is organised as follows: 
In the next section we discretise the free lagrangian with general
(first order) BKT for even scalars. In Section~3 we study its KK
decomposition and describe the corresponding results for odd scalars,
and for fermions and gauge bosons.  
Some conclusions are presented in the last section.

\section{DISCRETISED SCALAR MODELS}

As an example we write down a discretised version of the free
lagrangian for an even massless scalar $\phi$ with arbitrary BKT at one 
brane in a five dimensional model with the fifth coordinate compactified 
on an $S^1/Z_2$ orbifold
(compare with Eq. (2.1) of Ref. \cite{APS03b}):
\begin{eqnarray}
\mathcal{L}&=&
\sum_{i=0}^N  (1 - \frac{1}{2} (\delta_{i0} + \delta_{iN} ) 
+ \alpha \delta_{i0})
\partial_\mu \phi^\dagger_i \partial^\mu \phi_i \nonumber \\
&- & \frac{1}{s^2}  
\sum_{i=0}^{N-1} 
(\phi^\dagger_{i+1}-\phi^\dagger_{i})
(\phi_{i+1}-\phi_{i}) \nonumber \\
&+& 
 \frac{\beta}{s^2}  \Big[\phi^\dagger_0(\phi_1-\phi_0)
+\mathrm{h.c.}\Big]  \nonumber \\
&  - & 
\frac{\gamma}{s^2}  (\phi_1^\dagger -
  \phi_0^\dagger)(\phi_1-\phi_0) ,
\label{lagrangian}
\end{eqnarray}
where $s$ is the lattice spacing and $\alpha$, $\beta$ and~$\gamma$ 
are the dimensionless coefficients of the discretised BKT. The number 
of sites is $N+1$ and the boundary sites at $i=0,N$ play the role of 
branes. The spacing is the only dimensionful coefficient. The 
relation of these parameters with the parameters of the continuum 
theory is given by  
\begin{eqnarray} 
s=\frac{\pi R}{N}, \quad 
\frac{\alpha}{a}=
\frac{\beta}{b}=
\frac{\gamma}{c}=
\frac{N}{2\pi R},
\end{eqnarray}
where $R$ is the compactification radius and $a$, $b$ and $c$ are
defined in~\cite{APS03b}. We shall take $R$ constant, such that the
continuum limit $s\rightarrow 0$ corresponds to $N \rightarrow
\infty$.  

In order to define the BKT correctly, we consider a discrete version
of the theory with BKT compactified on a circle, compatible with the
parity symmetry $i\leftrightarrow -i$, and then identify
$\phi_{-i}\sim \phi_i$. In this way we can separate modifications of
the generalised mass matrix (which incorporates the discrete five
dimensional derivatives) near the branes due to BKT from the ones
which impose Neumann boundary conditions in the fundamental region for
even fields in the absence of BKT. More details are given
in~\cite{aliphatic}. There are, of course, discretisation ambiguities
in the definition of the discrete delta functions. Usually, one
expects that the continuum limit is independent of discretisation
details, but keeping in mind the singular behaviour associated to BKT
in the continuum, this is an issue to be checked. We have written
above the minimal versions of the discrete BKT compatible with the
orbifold parity, but have checked that the results, apart from small
side-effects in some cases, remain the same for
discrete delta functions involving a small number of sites in the
vicinity of the branes. 

In matrix notation, Eq.~(\ref{lagrangian}) reads (sum over $i,j$ is understood)
\begin{equation}
\mathcal{L}=\partial_\mu \phi_i^\dagger \mathcal{K}_{ij} \partial^\mu
\phi_j - \phi_i^\dagger
\mathcal{M}^2_{ij} \phi_j,
\label{matrixaction}
\end{equation}
where the 
$(N+1)\times (N+1)$
kinetic matrix $\mathcal{K}$ is
$$
\mathrm{diag}(\frac{1}{2}+\alpha,1,\ldots,1,\frac{1}{2})
$$
and the 
$(N+1)\times (N+1)$
mass matrix $s^2 \mathcal{M}^2$ is
$$ \begin{pmatrix}
1+2\beta+\gamma 
& -1-\beta-\gamma & 0 & \dots & 0 & 0 \\
-1-\beta-\gamma & 2+\gamma & -1 & \dots & 0 & 0 \\
0 & -1 & 2 & \dots & 0 & 0 \\
\vdots & \vdots & \vdots & \ddots & \vdots & \vdots \\
0 & 0 & 0 & \dots & 2 & -1 \\
0 & 0 & 0 & \dots & -1 & 1 \\
\end{pmatrix} . 
$$ 
One can discretise the lagrangians
for odd scalars, and for fermions and gauge bosons in a similar way 
(Eqs. (2.25) and (2.14) in Ref. \cite{APS03b}, respectively).

Solving this theory amounts to diagonalising the mass matrix 
after canonically normalising the kinetic term. 
Then  
\begin{equation}
\mathcal{L}=
\partial_\mu\varphi_i^\dagger \partial^\mu \varphi_i 
- m_i^2  \varphi_i^\dagger \varphi_i ,
\label{diagonalaction}
\end{equation}
with KK modes $\varphi_i$ and masses $m_i$ given by
\begin{equation}
\mathcal{K}^{\frac{1}{2}}_{ij} \phi_j = a_{in} \varphi_n
\label{Keq}
\end{equation}
and
\begin{equation}
\mathcal{M}^2_{ij} = 
\mathcal{K}^{\frac{1}{2}}_{ik} a_{kn} m^2_{n} a_{nl} 
\mathcal{K}^{\frac{1}{2}}_{lj} .
\label{diagonalisation}
\end{equation}
The orthogonal matrix $a_{in}$ is the discrete analogue of the KK wave
functions. We have assumed that $\mathcal{K}$ is positive definite, in
agreement with the usual positivity requirements of quantum field
theory. Then, $\mathcal{K}^{\frac{1}{2}}$ is real and
Eq.~(\ref{diagonalaction}) follows directly from
Eqs.~(\ref{matrixaction},\ref{Keq},\ref{diagonalisation}). 
The masses and wave functions can be found 
analytically or numerically from Eq.~(\ref{diagonalisation}).
Although we shall not discuss gravity here, we note in passing that the
localised kinetic terms in DGP models~\cite{DGP} are of the $a$ type. 

\section{FIELD SPECTRA}

Let us now compare the results with the ones obtained with a thick
brane regularisation in Ref. \cite{APS03b}. We focus mainly on the  
relation between the corresponding large $N$ and thin brane
limits. Before doing that, we clarify one important point regarding
the continuum limit: 
The coefficients of the BKT in the continuum have dimensions of length
and are suppressed by one power of the spacing, which acts as the
inverse UV cutoff of the effective theory describing the finite-$N$
deconstructed theory at low energies. Note that the same applies, for
instance, to gauge coupling constants. 
Here we simply want to use the discrete theory as a regulator to
understand the classical behaviour of the lowest dimensional
lagrangian with BKT in the continuum. Then, in taking the continuum
limit, we need to keep the parameters $a$, $b$ and $c$ fixed. This
implies that the discrete parameters $\alpha$, $\beta$ and $\gamma$
are taken to be proportional to $N$. 

\subsection{Scalars}

For even scalars 
and $a$ type BKT $(b = c = 0)$ the results
for the leading term in the $1/N$ expansion 
(continuum limit), which can be easily obtained
analytically, are the same as for the thin brane regularisation 
\cite{APS03b}.
There is always a constant massless mode, and for  
large positive $a$ the non-zero masses approach $\frac{1}{2},
\frac{3}{2}, ...$, in units of $R$.  
For negative $a$ Eqs.~(\ref{diagonalaction},\ref{Keq},\ref{diagonalisation}) 
must be generalised to deal with ghosts, 
possibility which is not discussed here.

\begin{figure}[!htb]
\begin{center}
\epsfig{file=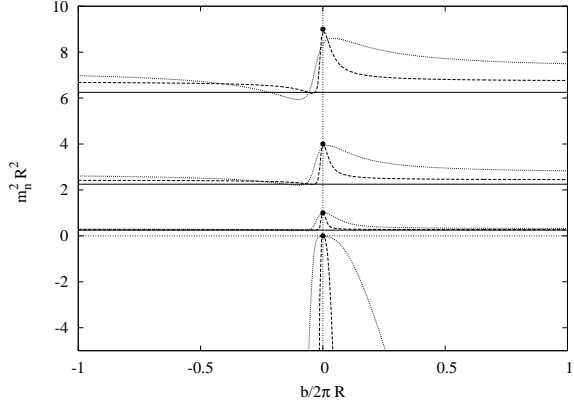,height=5.5cm}
\end{center}%
\caption{Squared masses of the lightest modes in units of $R^2$ as a function 
of $\frac{b}{2\pi R}$ in the range $[-1, 1]$ 
for $N$ = 10 (dots), 25 (dashes) and the thin brane limit 
(solid). The dots correspond to $b = 0$ 
and $m_n = \frac{n}{R}$.}
\label{fig:b}
\end{figure}

\begin{figure}[!htb]
\begin{center}
\epsfig{file=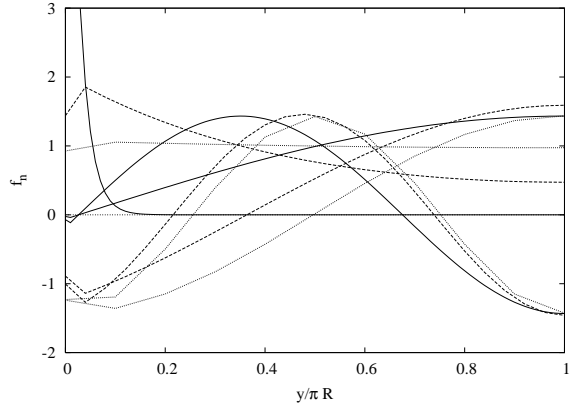,height=5.5cm}
\end{center}
\caption{Eigenfunctions of the lightest modes for $b = \frac{R}{10}$ and  
$N$ = 10 (dots), 25 (dashes) and 100 (solid). 
We plot $f_n(\frac{y}{\pi R} \equiv \frac{i s}{\pi R}) \equiv 
\sqrt{\frac{\pi R}{s}} a_{in}$ for $n = 0$ (tachyon), 1 and 2.}
\label{fig:b1}
\end{figure}

As in the thin brane limit of the thick brane regularisation, the large 
$N$ limit commutes with 
$a\rightarrow 0^+$. However, this is not so for $b$ as is apparent from 
Fig. \ref{fig:b}, where we draw the masses 
of the lightest KK modes for $N = 10, 25$, as well as 
for the continuum limit, as a function of $b (a, c = 0)$. 
For $b = 0$ the masses are $0, 1, 2 , ...$ (in units of $R$), 
but for a non-zero $b$, independently of how small it is, 
the zero mode becomes tachyonic and the massive modes decrease to
$\frac{1}{2}, \frac{3}{2}, ...$ for large $N$.  
The tachyonic mass squared tends to 
$-\infty$ for  
any non-zero value of $b$ for large $N$, and the corresponding eigenfunction 
to a normalizable function peaked only at the origin 
(see Fig. \ref{fig:b1}). 
The other eigenfunctions approach
$\sim \sin (\frac{2n+1}{2R}|y|)$.
(For small $b$ and $N$ they look like 
$\sim -\cos (\frac{n}{R}y)$.)
This implies that in the continuum limit all the bulk modes but the
tachyon are decoupled from the brane fields. These modes are also
decoupled from the tachyon, which has support only at the origin.  
All these results are qualitatively 
independent of the sign of $b$, in contrast with the results in the
thick brane regularisation. There, the same behaviour is obtained for
$b\geq 0$ (although the presence of a tachyon was overlooked in
\cite{APS03b}), but 
for negative $b$ the eigenfunctions 
diverge at some point near $y = 0$. One
can define a {\it principal value} (PV) sort of limit adding a tiny 
imaginary part to $b$ in the equation of motion, and taking it to zero 
after performing the thin brane limit. This is presumably equivalent to 
excluding a symmetric 
interval around the divergent points, and taking the thin brane 
limit first and the interval to zero afterwards; this explains the 
notation PV. We have found that with this prescription the large $N$ limit 
result in 
Fig. \ref{fig:b} is recovered also for negative $b$. Therefore the 
singularity near the origin may be an artifact of the thick brane 
regularisation.
It is worth emphasizing at this point that the discrete regularisation 
is easier to handle as it does not have the convergence and 
numerical problems of the thick brane regularisation. 

\begin{figure}[htb]
%\framebox[55mm]{\rule[-21mm]{0mm}{43mm}}
\begin{center}
\epsfig{file=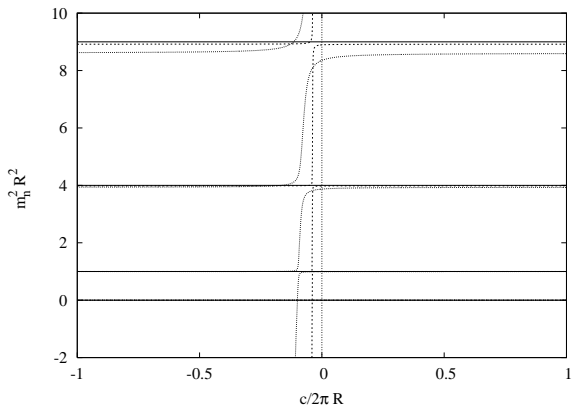,height=5.5cm}
\end{center}
\caption{Squared masses of the lightest modes 
in units of $R^2$ as a function 
of $\frac{c}{2\pi R}$ in the range $[-1, 1]$ 
for $N$ = 10 (dots), 25 (dashes) and the thin brane limit 
(solid).}
\label{fig:c}
\end{figure}

In Fig. \ref{fig:c} we plot the masses 
of the lightest KK modes for $N = 10, 25$, as well as 
for the continuum limit, as a function of $c (a, b = 0)$. 
For $c < 0$ the zero mode is not affected but the first 
massive mode becomes tachyonic in a similar way as the 
zero mode does for $b \neq 0$. The other massive modes decrease 
their masses in one unit. Using the PV prescription, the thin brane 
limit of the continuum theory coincides with the large $N$ limit of the 
discretised theory. For non-negative $c$ both limits give no dependence on $c$.

Odd scalars are not sensitive to BKT \cite{APS03b}.

\subsection{Fermions}

Let us now deal with fermions. From the four-dimensional point 
of view they are Dirac spinors with two chiral
components:
$\Psi=\Psi_L+\Psi_R$, $\gamma^5 \Psi_{L,R}=\mp \Psi_{L,R}$.
The invariance
of the bulk kinetic term requires that the left-handed (LH) and
right-handed (RH) components have opposite $Z_2$ parities.
We assume without loss of generality that the LH (RH) component 
is even (odd), and consider the general BKT $a^L$, $a^R$, $b$ and $c$
in Eq. (2.25) of Ref. \cite{APS03b}. 
In order to avoid the (in)famous doubling problem we include a Wilson
term.
More details are given in~\cite{aliphatic}. 
Here we directly
discuss the results~\footnote{The KK reduction proceeds as in the
scalar case, but  
in this case one must diagonalise the matrix 
$\mathcal{M}\mathcal{M}^\dagger$, where $\mathcal{M}$ is 
the discretised fermion mass matrix.}.

As in the thick brane regularisation, there is always a LH chiral 
zero mode. The solutions for $a^L (a^R, b, c = 0)$ 
coincide with those for scalars of the same parity. 
$a^R$ is irrelevant when $a^L = 0$, and when both are 
non-zero the limits $a^L \rightarrow 0$, $a^R \rightarrow 0$ and 
$N\rightarrow \infty$ commute.
For non-zero $b$ the behaviour of fermions is quite different from the one of  
scalars. In the continuum limit the masses and the eigenfunctions away from 
the brane are the same as for vanishing 
$b$. At the brane, the even wave functions are discontinuous 
and zero. This behaviour is independent of the sign of $b$. Again, one must 
use the PV prescription in the thick brane case to reproduce it for negative 
$b$. 

It is worth comparing now the scalar and fermion 
spectra for non-zero $b$. Although in both cases the bulk tends 
to decouple from the brane, 
in the scalar case there is a tachyon of infinite mass, 
localised at the brane, which 
decouples from the KK tower because all the other 
eigenfunctions vanish at the origin. In the 
fermionic case there are, of course, no tachyons but the even functions 
become discontinuous at the origin, where they vanish.
Moreover, scalar masses are shifted down to half-integer 
values, whereas fermion masses remain unchanged. In both
cases there is no explicit dependence on the $b$ value, 
as long as it is non-zero, and the $b \rightarrow 0$ limit is always 
singular. 
Imposing supersymmetry, which links scalars to fermions, 
one gets rid of tachyons. This is technically so because the 
supersymmetric scalar lagrangian includes $b^2 \delta^2$ 
terms which drastically modify the scalar behaviour, 
resulting in the same solutions as for fermions \cite{APS03b}. 
%Deconstruction of BKT in supersymmetric theories could be performed
%following the formalism in~\cite{CsakiSUSYdecon}, but this is beyond
%the scope of this work. 

For fermions, the only effect of a non-zero $c$ in the continuum
limit, independently of its sign, is to cancel the effect of
$a^R$. This holds in the thin brane limit of the thick brane
regularisation (with the PV prescription) as well. This behaviour is
also singular for $c\rightarrow 0$ if $a^R \neq 0$ and $a^L \neq
0$. The parameter $b$ is always determinant, since for $b \neq 0$ the
bulk fields decouple from the brane and the other  
BKT become irrelevant.

\subsection{Gauge bosons}

The situation for gauge bosons is a particular case of the one for scalars. 
The only new feature is that gauge invariance requires $b=0$ for the 
four-dimensional components of the gauge field, and $b=c=0$ for the 
fifth-dimensional one.

\section{CONCLUSIONS}

We have used a discrete regularisation to analyse 
the behaviour of BKT for bulk fields in brane models 
with extra dimensions. 
This regularisation has proven to be very efficient as compared with
regularisations of BKT in the continuum used before, and has allowed
us to identify some tachyonic modes which had evaded our previous
numerical computations. Moreover, discretisation is free of problems
which are present in the thick brane regularisation when some
parameters take negative values. This has led us to propose a PV
prescription for the thick brane, such that the results with both
regularisations agree in all cases. 
Except for positive $a$ type terms, the BKT we have studied 
show a singular behaviour when they approach $0$.  
It is clear that theories without
``dangerous'' BKT are distinguished, since they do not present these
problems. In principle, supersymmetric models with $b=c$ chosen to be
zero at tree level are stable under quantum corrections and belong to
this class. Putting $b=c=0$ could be justified in the context of
string theory. More generally, one can construct critical theories by
including a tower of higher-dimensional operators such that the
singular behaviour is smoothed down, as proposed in
Ref.~\cite{APS03b}. This is equivalent to a classical renormalisation
of the theory. 
  
Even in the noncritical case, it is plaussible that the microscopic theory 
takes care of the singular behaviour via some mechanism which introduces a 
new scale (possibly related to the cutoff scale controlling the dimensionful 
coefficients of the BKT). In fact, both the thick brane and the lattice 
regularisations suggest that this is unavoidable, as they both introduce a 
length parameter: the brane width and the spacing, respectively. From a low 
energy perspective, in organising the perturbative series one should take 
into account that there are two scales at play, and make sure that the 
effective lagrangian to be used describes the actual physics in the regime 
of interest.

\end{document}